\documentclass[twocolumn]{autart}

\pdfoutput=1

\usepackage{graphicx}          

\usepackage{acronym}
\usepackage{color}
\usepackage{amsmath} 
\usepackage{amssymb} 
\usepackage{subfigure}
\usepackage{enumitem}  
\usepackage{algorithmic}
\usepackage[process=auto]{pstool}
\usepackage{multirow} 
\usepackage{url} 
\newtheorem{remark}{Remark}

\begin{document}

\acrodef{FDIR}{Fault Detection, Isolation and Reconstruction}
\acrodef{FDIE}{Fault Detection, Isolation and Estimation}
\acrodef{FDD}{Fault Detection and Diagnosis}
\acrodef{FTC}{Fault Tolerant Control}
\acrodef{FDI}{Fault Detection and Isolation}
\acrodef{IMU}{Inertial Measurement Unit}
\acrodef{ADS}{Air Data Sensors}
\acrodef{GPS}{Global Positioning Systems}
\acrodef{SMO}{Sliding Mode Observers}
\acrodef{DOS}{Dedicated Observer Scheme}
\acrodef{KF}{Kalman Filter}
\acrodef{EKF}{Extended Kalman Filter}
\acrodef{IEKF}{Iterated Extended Kalman Filter}
\acrodef{UKF}{Unscented Kalman Filter}
\acrodef{AUKF}{Augmented Unscented Kalman Filter}
\acrodef{MM}{multiple model}
\acrodef{DMAE}{Double-Model Adaptive Estimation}
\acrodef{DMAE-NSR}{Double-Model Adaptive Estimation-No Selective Reinitialization}
\acrodef{MMAE}{Multiple-Model Adaptive Estimation}
\acrodef{IMM}{Interacting Multiple-Model}
\acrodef{AMMAE}{Augmented Multiple-Model Adaptive Estimation}
\acrodef{SRMMAE}{Selective-Reinitialisation Multiple Model Adaptive Estimation}
\acrodef{FF}{Fault Free}
\acrodef{SR}{Selective-Reinitialization}
\acrodef{ADDSAFE}{Advanced Fault Diagnosis for Sustainable Flight Guidance and Control}

\acrodef{RTSKF}{Recursive Three-Step Kalman Filter}
\acrodef{ATSKF}{Adaptive Three-Step Kalman Filter}
\acrodef{RMSE}{root mean square error}
\acrodef{OTSKF}{Optimal Two-Stage Kalman Filter}
\acrodef{AFKF}{Adaptive Fading Kalman Filter}
\acrodef{AKF}{Augmented Kalman Filter}
\acrodef{R4SKF}{Recursive Four-Step Kalman Filter}
\acrodef{A2KF}{Adaptive Augmented Kalman Filter}

\begin{frontmatter}

\title{Simultaneous State and Unknown Input Estimation for Continuous-discrete Stochastic Systems\thanksref{footnoteinfo}}

\thanks[footnoteinfo]{This work was supported by the Hong Kong University Grants Committee (UGC) Fund under Grant 1-BE24.
This paper was not presented at any IFAC meeting. }

\author[cs]{Peng Lu\thanksref{footnoteinfo2}}\ead{peng.lu@polyu.edu.hk}   

\address[cs]{The Hong Kong Polytechnic University, HKSAR, China}

\thanks[footnoteinfo2]{The author is with the Adaptive Robotic Controls Lab (ArcLab), the Hong Kong Polytechnic University. Website: \url{https://www.polyu.edu.hk/researchgrp/arclab/arclab.html}
Tel. +852 34008065. Fax +852 2725 4922. }


\begin{abstract}                          
This paper considers the simultaneous state and unknown input estimation for continuous-discrete stochastic systems. Two types of approaches (with and without modeling of unknown inputs) which can address this issue are investigated. A novel continuous recursive four-step Kalman filter is proposed and its asymptotic stability condition is established. A novel one-step unknown input Kalman filter is proposed and has guaranteed stability when the number of unknown inputs is equal to that of the measurements. The design of unknown input Kalman filters and observers is unified. Furthermore, an adaptive augmented Kalman filter which requires the modeling of unknown inputs is introduced. The estimation error covariance of the recursive four-step Kalman filter and the adaptive augmented Kalman filter is analyzed and compared. Finally, simulation results demonstrate the effectiveness of the proposed approaches.
\end{abstract}

\end{frontmatter}


\section{Introduction}
\label{s:1}
Simultaneous state and unknown input estimation has been considered for a few decades. Generally, there are two types of Kalman filters which can address the state and unknown input estimation: one augments the unknown inputs as states whereas the other one does not. The one does not augment unknown inputs is usually called unknown input Kalman filters whereas the other one is called augmented Kalman filters. 

Due to the fact that unknown inputs are difficult to model, unknown input Kalman filters have received a lot of attention and many approaches are presented \cite{Kitandis1987,Darouach1997,Hou1998,Hsieh2000,Darouach2003,Gillijns2007c,Gillijns2007,Cheng2009,Hsieh2009}.  The principle of these approaches are the same, which is to decouple unknown inputs from the states such that the state estimation is not affected by the unknown inputs. Consequently, most of them are equivalent in terms of either state estimation or unknown input estimation whereas the differences lie in how unbiased state estimates or unknown input estimates are derived. Some recent works can be found in \cite{Yong2016,Kim2020,Hsieh2020}. However, all these approaches are proposed for discrete-time systems. Yong \cite{Yong2017} proposed an unknown input Kalman filter for continuous stochastic systems. However, many assumptions are made which include the process and measurement noises and also the input and unknown inputs. These assumptions are necessary to derive the unknown input Kalman filter.

In contrast, the other type of approach, augmented Kalman filters, have received little attention for state and unknown input estimation. This is mostly caused by the fact that unknown inputs have unpredictable behaviors and it is challenging to model its dynamics. A common practice is to model unknown inputs as a stochastic process driven by a white noise. However, the covariance matrix of the white noise is difficult to determine. Another drawback of augmented Kalman filter is its computational load. To address that, Friedland proposed a two-stage Kalman filter which splits the augmented Kalman filter into a bias-free filter and a bias filter. It was extended by Hsieh and Chen \cite{Hsieh1999} to deal with time-varying biases. Lu et al. \cite{Lu2015e} extended the two-stage Kalman filter to deal with nonlinear systems. It should be noted that due to fast development of the computational power of computers nowadays, the computational load introduced by the augmented states is not of great concern anymore. Furthermore, Lu et al. \cite{Lu2016} proposed an adaptive two-stage Kalman filter which can estimate the covariance matrix of the unknown inputs. Therefore, the two issues regarding the augmented Kalman filter can be considered as solved to a certain extent. There exist more complex forms of Kalman filters,  which are based on multiple models or double models, to address simultaneous state and unknown input estimation \cite{Ducard2008,Lu2015g,Lu2015,Lu2016a}. However, they can be all generalized as augmented Kalman filters.

This paper will consider the simultaneous state and unknown input estimation using the two types of approaches. The main contributions of this paper are as follows:
\begin{enumerate}
\item A novel \ac{R4SKF} for continuous stochastic systems is proposed and the stability is analyzed. 
\item A one-step unknown input Kalman filter with guaranteed stability is proposed for the case when the number of unknown inputs is the same as that of the measurements. It is theoretically proved that the optimal Kalman gain of unknown input Kalman filters has no effect on the final state estimates.
\item It is proved that for addressing continuous systems, unknown input Kalman filters are equivalent to unknown input observers if properly designed.
\item The estimation performance of the proposed \ac{R4SKF} and an \ac{A2KF} is analyzed theoretically.
\end{enumerate}

The structure of this paper is as follows: the problem is defined in Sec.~\ref{s:2}. The novel \ac{R4SKF} for continuous stochastic systems is proposed in Sec.~\ref{s:3}. In Sec.~\ref{s:4}, a simplified unknown input Kalman filter is proposed when the number of unknown inputs are the same as that of the measurements. Sec.~\ref{s:5} presents the relationship between unknown input Kalman filters and unknown input observers when addressing continuous stochastic systems. The \ac{A2KF} and the theoretical analysis of its estimation performance comparison with the \ac{R4SKF} is given in Sec.~\ref{s:6}. Sec.~\ref{s:7} compares the two approaches using simulated examples and conclusions are presented in Sec.~\ref{s:8}.

\section{Problem formulation}
\label{s:2}
Consider the following time-varying continuous-discrete system with unknown inputs:
\begin{align}
\label{e:x_dot}
\dot{x}(t) & = A(t) x(t) + B(t) u(t) + E(t) d(t) + G(t) w(t) \\
y_{k} & = C_{k} x_{k}  + \nu_{k}
\label{e:y}
\end{align}
where $x_k=x(t_k)$, $x_k\in \mathbb{R}^{n_x}$ represents the state vector, ${y}_k\in \mathbb{R}^{n_y}$ is the measurement/output vector, ${d}(t) \in \mathbb{R}^{n_d}$ is the unknown input vector. It can represent disturbances or sensor faults. $A(t)$, $B(t)$, $E(t)$, $G(t)$ and $C_k$ are known matrices with appropriate dimensions. ${w}(t) \sim N(0,Q(t))$ and ${v}_k\sim N(0,R_k)$ are the process noise and measurement noise vector, respectively. $u(t)\in \mathbb{R}^{n_u}$ represents the known inputs. 

Note that most physical systems can be represented using this continuous-discrete system since measurements are usually obtained at a discrete time step.

It is assumed that the system is observable. Moreover, it is assumed that no prior knowledge about the dynamics of $d_k$ is available. $d_k$ can be any type of signal. Without losing generality, we follow the assumption that rank $CE=$ rank $E$. This assumption ensures that an unknown input Kalman filter can be designed.


\begin{remark}
The model described by Eqs.~\eqref{e:x_dot} and \eqref{e:y} represents the problem of disturbance estimation or input/actuator fault estimation.
\end{remark}

\section{A Recursive Four-Step Continuous Kalman Filter}
\label{s:3}
This section proposes the novel \ac{R4SKF} with the solutions to each step. The stability of the proposed filter is analyzed. Furthermore, the approach is extended to nonlinear systems.
\subsection{Sketch of the \ac{R4SKF}}
\label{s:31}

We propose the following recursive four-step Kalman filter:
\begin{itemize}
\item[] Step 1: Prediction without unknown inputs\\
Solve $x^*_{k|k-1}=x^*(t_k|t_{k-1})$ using
\begin{align}
\label{e:x_pred nod}
\dot{x}^*(t|t_{k-1}) & = A(t) \hat{x}(t|t_{k-1}) + B(t) u(t)
\end{align}
with $\hat{x}(t_{k-1}|t_{k-1})=\hat{x}_{k-1|k-1}$.

\item[] Step 2: Unknown input estimation\\
Solve $\hat{d}_{k-1}=\hat{d}(t_{k-1})$ using 
\begin{align}
\label{e:d_est}
\hat{d}_{k-1} = f_d(y_k - C_k x^*_{k|k-1} )
\end{align}
where $f_d$ is a function which will be discussed later. The aim of this function is to solve $\hat{d}_{k-1}$ from $y_k - C_k x^*_{k|k-1} $.

\item[] Step 3: Prediction with unknown inputs\\
Solve $\hat{x}_{k|k-1}=\hat{x}(t_k|t_{k-1})$ using
\begin{align}
\label{e:x_pred d}
\dot{\hat{x}}(t|t_{k-1}) & = A(t) \hat{x}(t|t_{k-1})  + B(t) u(t) + E(t) \hat{d}(t)
\end{align}
with $\hat{x}(t_{k-1}|t_{k-1})=\hat{x}_{k-1|k-1}$ and $\hat{d}(t_{k-1})=\hat{d}_{k-1}$.

\item[] Step 4: Measurement update\\
\begin{align}
\label{e:x_est}
\hat{x}_{k|k} = \hat{x}_{k|k-1} + K_k (y_k - C_k \hat{x}_{k|k-1} )
\end{align}
\end{itemize}
For readability, ${x}^*(t|t_{k-1})$ and $\hat{x}(t|t_{k-1})$ will be replaced by ${x}^*(t)$ and $\hat{x}(t)$, respectively.  

Gillijns and De Moor \cite{Gillijns2007} proposed a similar framework for unknown input estimation. However, their method is only applicable to discrete-time systems. The solutions to each step of the \ac{R4SKF} is presented in the following.

\subsection{Prediction without unknown inputs}
\label{s:32}
Let $\hat{x}(t_{k-1})=\hat{x}_{k-1|k-1}$ be unbiased, then we can predict the state by assuming there are no unknown inputs as follows:
\begin{align}
\dot{x}^*(t) & = A(t) \hat{x}(t) 
\end{align}
Since the system is linear, the following solution is obtained:
\begin{align}
\label{e:x_star_tk}
x^*(t_k) = \Phi(t_k,t_{k-1}) \hat{x}(t_{k-1}) + \int_{t_{k-1}}^{t_k} \Phi(t_k,\tau) B(\tau) u(\tau) d\tau
\end{align}
where $\Phi(t_k,t_{k-1})=e^{A(t_k-t_{k-1})}=e^{A\Delta t}$ is the transition matrix. Similarly, $\Phi(t_k,\tau)=e^{A(t_k-\tau)}$. Assuming piecewise constant inputs over a sampling period, \eqref{e:x_star_tk} is equal to:
\begin{align}
\label{e:x_star_tk u}
x^*(t_k) = \Phi(t_k,t_{k-1}) \hat{x}(t_{k-1}) + u(t_{k-1})\int_{t_{k-1}}^{t_k} \Phi(t_k,\tau) B(\tau)  d\tau
\end{align}

If higher order terms are neglected, the above equation can be further simplified as follows:
\begin{align}
\label{e:x_star_tk appro}
x^*(t_k) &= [I+A(t_{k-1})\Delta t] \hat{x}(t_{k-1}) + [B(t_{k-1})\Delta t] u(t_{k-1}) \\
&:= A_d(t_{k-1}) \hat{x}(t_{k-1}) + B_d(t_{k-1}) u(t_{k-1})
\end{align}

Consequently, the state prediction without effects of unknown inputs, denoted by $x^*_{k|k-1}=x^*(t_k)$, is obtained. The method presented here is only valid for linear systems. The solution for nonlinear systems will be introduced in Sec~\ref{s:35}.

%

\subsection{Unknown input estimation}
\label{s:33}
The measurement vector $y_k$ can be written as follows:
\begin{align}
\label{e:yk int}
y_k =& C_k [\Phi(t_k,t_{k-1}) {x}(t_{k-1}) + \int_{t_{k-1}}^{t_k} \Phi(t_k,\tau) ( B(\tau)u(\tau) \nonumber \\
&+ E(\tau)d(\tau) + G(\tau)w(\tau) ) d\tau ]
\end{align}

Define $\gamma^*_{k}$ as
\begin{align}
\label{e:gamma_d}
\gamma^*_{k} = y_k - C_k x^*_{k|k-1}
\end{align}
Then $\hat{d}(t_{k-1})$ is solved using the following:
\begin{align}
\int_{t_{k-1}}^{t_k} \Phi(t_k,\tau) E(\tau)d(\tau) d\tau= \gamma^*_{k}.
\end{align}
The solution is denoted as $\hat{d}(t_{k-1}) = f_d(\gamma^*_{k})$.

By neglecting higher order terms and substituting \eqref{e:yk int} into \eqref{e:gamma_d}, it follows:
\begin{align}
\label{e:gamma_star}
\gamma^*_{k} =& C_k A_d(t_{k-1}) e_{k-1|k-1} + C_k E_d(t_{k-1}) d(t_{k-1}) \nonumber \\
& + C_k G_d(t_{k-1})w(t_{k-1}) + v_k
\end{align}
where $e_{k-1|k-1} =x(t_{k-1})- \hat{x}_{k-1|k-1}$. Since rank $(CE)$=rank $E$, the unknown inputs $d(t_{k-1})$ can be estimated unbiasedly using the following:
\begin{align}
\label{e:d_est_exact}
\hat{d}(t_{k-1}) &= (C_k E_d(t_{k-1}))^{+} \gamma^*_{k} \nonumber \\
& := F_d(t_{k-1}) \gamma^*_{k} 
\end{align}
where $(C_k E_d(t_{k-1}))^{+}$ denotes the Moore–Penrose inverse of $C_k E_d(t_{k-1})$ with $E_d(t_{k-1}) = E(t_{k-1})\Delta t$. Consequently, $f_d(\gamma^*_{k})$ is reduced to $F_d(t_{k-1}) \gamma^*_{k}$.

\subsection{Prediction with unknown inputs}
\label{s:34}
Once unbiased estimates of $\hat{d}(t_{k-1})$ is obtained, the full prediction, which contains the unknown inputs, can be performed as follows:
\begin{align}
\dot{\hat{x}}(t) & = A(t) \hat{x}(t)  + B(t) u(t) + E(t) \hat{d}(t)
\end{align}
with initial condition give by $\hat{x}_{k-1|k-1}$ and $\hat{d}(t_{k-1})$. The solution to this is the same as in Sec~\ref{s:32}. Up until now, the following theorem is obtained.

\begin{thm}
\label{thm:unbiased estimator}
Let $\hat{x}_{k-1|k-1}$ be unbiased, the estimator denoted by \eqref{e:x_pred nod}-\eqref{e:x_pred d} is unbiased. 
\end{thm}
\begin{pf}
Define the prediction error $e_{k|k-1}$ as follows:
\begin{align}
e_{k|k-1} =& x_k - \hat{x}_{k|k-1} \\
\label{e:ek_k-1}
=& A_d(t_{k-1}) e_{k-1|k-1} + E_d(t_{k-1}) e^d(t_{k-1}) \nonumber \\
& + G_d(t_{k-1}) w(t_{k-1}) 
\end{align} 
where $e^d(t_{k-1}) = d(t_{k-1}) - \hat{d}(t_{k-1})$ and $G_d(t_{k-1})=G(t_{k-1})\Delta t$.

As unbiased $\hat{d}(t_{k-1})$ is obtained by unbiased $ \hat{x}_{k-1|k-1}$, the estimator denoted by \eqref{e:x_pred nod}-\eqref{e:x_pred d} is unbiased as long as $\hat{x}_{k-1|k-1}$ is unbiased.
\rlap{$\qquad \Box$}
\end{pf}

The stability of this estimator is given by the following theorem:
\begin{thm}
\label{thm:unbiased estimator stability}
The estimator denoted by \eqref{e:x_pred nod}-\eqref{e:x_pred d} is stable if and only if all the eigenvalues of $( I-E_d(t_{k-1})F_d(t_{k-1})C_k ) A_d(t_{k-1})$ are within the unit circle. 
\end{thm}
\begin{pf}
The dynamics of the estimation error are
\begin{align}
{e}_{k|k-1} =& A_d(t_{k-1}) e_{k-1|k-1} + E_d(t_{k-1}) e^d(t_{k-1}) \nonumber \\
&+ G_d(t_{k-1}) w(t_{k-1})
\end{align}
%
To analyze the stability, we still need to compute $e^d(t_{k-1})$. Combing \eqref{e:gamma_star} and \eqref{e:d_est_exact}, $e^d(t_{k-1})$ is obtained as follows:
\begin{align}
\label{e:ed}
e^d(t_{k-1}) = & F_d(t_{k-1}) (-C_k A_d(t_{k-1}) e_{k-1|k-1} \nonumber \\
& - C_k G_d(t_{k-1})w(t_{k-1}) -v_k )
\end{align}

Consequently, the prediction error in \eqref{e:ek_k-1} can be further expressed as follows:
\begin{align}
\label{e:ek_k-1_nod}
e_{k|k-1} =& \bar{A}_{k-1}  e_{k-1|k-1} + \bar{G}_{k-1} w(t_{k-1}) + \bar{D}_k v_k
\end{align}
where
\begin{align}
\bar{A}_{k-1} &= ( I-E_d(t_{k-1})F_d(t_{k-1})C_k ) A_d(t_{k-1}) \\
\bar{G}_{k-1} &= (I-E_d(t_{k-1})F_d(t_{k-1})C_k) G_d(t_{k-1}) \\
\bar{D}_k &= - E_d(t_{k-1})F_d(t_{k-1})
\end{align}
Therefore, to guarantee the stability, all the eigenvalues of $\bar{A}_k$ should lie within the unit circle.
\rlap{$\qquad \Box$}
\end{pf}


\subsection{Measurement update}
\label{s:35}
Theorem~\ref{thm:unbiased estimator} is based on the assumption that $\hat{x}(t_{k-1})$ is unbiased. By performing measurement update, this assumption can be removed.  
The measurement update, denoted as in \eqref{e:x_est}, can be performed as follows \cite{Kitandis1987,Darouach1997,Gillijns2007c}:
\begin{align}
P^*_{k|k-1} =& A_d(t_{k-1}) P^*_{k-1|k-1} A_d(t_{k-1})^T \nonumber \\
& + G(t_{k-1}) Q(t_{k-1}) G(t_{k-1})^T \Delta t \\
K_k =& P^*_{k|k-1} C_k^T (C_k P^*_{k|k-1} C_k^T + R_k)^{-1} \\
L_k =& K_k + (I-K_k C_k) E_d(t_{k-1}) F_d(t_{k-1}) \\
P^*_{k|k} =&  (I-L_k C_k)P^*_{k|k-1} (I-L_k C_k)^T + L_k R_k L_k^T
\end{align}
The optimal Kalman gain is not unique for discrete-time systems \cite{Gillijns2007c},which also holds for continuous systems. A general form of the optimal gain matrix is given in \cite{Darouach1997,Gillijns2007c}. The effect of the Kalman gain will be discussed further in Sec.~\ref{s:4}.

The stability of this filter is given in the following theorem.
\begin{thm}
\label{thm:R4sKF stability}
The estimator denoted by \eqref{e:x_pred nod}-\eqref{e:x_est} is stable if and only if all the eigenvalues of $(I - K_k C_k) \bar{A}_{k-1}$ is within the unit circle.
\end{thm}
\begin{pf}
The estimation \eqref{e:x_est} can be rewritten as
\begin{align}
\hat{x}_{k|k-1} = (I-K_k C_k) \hat{x}_{k|k-1} + K_k y_k
\end{align}
Therefore, the error dynamics of the estimation are:
\begin{align}
\label{e:ek_k}
e_{k|k} =& (I-K_k C_k) e_{k|k-1} - K_k v_k
\end{align}
Substituting \eqref{e:ek_k-1_nod} into the above equation, yields
\begin{align}
e_{k|k} &= \tilde{A}_{k-1} e_{k-1|k-1} + \tilde{G}_{k-1} w(t_{k-1} ) + \tilde{D}_k v_k
\end{align}
where
\begin{align}
\label{e:A tilde}
\tilde{A}_{k-1} &= (I - K_k C_k) \bar{A}_{k-1} \\
\label{e:G tilde}
\tilde{G}_{k-1} &= (I - K_k C_k) \bar{G}_{k-1} \\
\label{e:D tilde}
\tilde{D}_k &= (I - K_k C_k) \bar{D}_k -K_k 
\end{align}
This completes the proof.
\rlap{$\qquad \Box$}
\end{pf}

\subsection{Extension to nonlinear systems}
\label{s:33}
We can readily extend the above properties to the following continuous-discrete time systems:
\begin{align}
\label{e:x_dot_non}
\dot{x}(t) & = f(x(t),u(t),t) + E(t) d(t) + G(t) w(t) \\
y_k & = h(x_k) +  \nu_k
\label{e:y_dis_non}
\end{align}

For nonlinear systems, the transition matrix can not be computed. The prediction of \eqref{e:x_dot_non} can be performed by
\begin{align}
x(t_k) &= x(t_{k-1}) + \int_{t_{k-1}}^{t_k} [f(x(\tau),u(\tau),\tau) + E(\tau) d(\tau) ] d\tau
\end{align}

The integral part can be solved using the Runge-Kutta algorithm. For simplicity, they can also be solved using Euler integration. The error covariance matrix $P(t_k)$ can be computed from
\begin{align}
\dot{P}(t) = F(t) P(t) + P(t) F(t)^T + G(t) Q(t) G(t)^T
\end{align}
where $F(t)$ is the linearized matrix of $f(x(t),u(t),t)$ with respect to $x(t)$. The Runge-Kutta algorithm can also be used. If Euler integration is used, we recommend the following to reduce the effects of neglecting higher order terms:
\begin{align}
P(t_k) =& P(t_{k-1}) +  [F(t_{k-1}) P(t_{k-1}) + P(t_{k-1}) F(t_{k-1})^T \nonumber \\
& + F(t_{k-1}) P(t_{k-1}) F(t_{k-1})^T \Delta t \nonumber \\
& + G(t_{k-1}) Q(t_{k-1}) G(t_{k-1})^T] \Delta t
\end{align}
Note that the above technique still uses the linearized matrix of $f(x(t),u(t),t)$. To avoid that, the reader is referred to the continuous-discrete form of the Unscented Kalman filter \cite{Sarkka2007}.

\section{A one-step Kalman filter}
\label{s:4}
In this section, we consider a special case when $n_x=n_y=n_d$. This is an extreme case when the number of measurements is equal to that of the unknown inputs. However, this case is common in disturbance or input fault detection problems.
It will be demonstrated in the following theorem that a simpler estimator can be designed for this special case. 

\begin{thm}
\label{thm:one-step}
Provided that $n_x=n_y=n_d$, unbiased state estimate can be obtained using the following one-step Kalman filter:
\begin{align}
\label{e:x_est_one_step}
\hat{x}_{k|k} = C_k^{-1} y_k
\end{align}
\end{thm}

\begin{pf}
The estimator \eqref{e:x_est} can be rewritten into the following:
\begin{align}
\label{e:x_est_special}
\hat{x}_{k|k} =& {x}^*_{k|k-1} + E_d(t_{k-1})\hat{d}(t_{k-1}) \nonumber \\
& + K_k (y_k - C_k {x}^*_{k|k-1} - C_k E_d(t_{k-1})\hat{d}(t_{k-1})) \\
=& {x}^*_{k|k-1} + K_k (y_k - C_k {x}^*_{k|k-1}) \nonumber \\
& + [ E_d(t_{k-1})- K_k C_k E_d(t_{k-1})] \hat{d}(t_{k-1})
\end{align}

Since rank $C_k E(t_{k-1})$ = rank $E(t_{k-1})=n_d$, both $C_k$ and $E_d(t_{k-1})$ are invertible.
Therefore, it follows
\begin{align}
F_d(t_{k-1})=E_d^{-1}(t_{k-1}) C_k^{-1}.
\end{align}

According to \eqref{e:d_est_exact}, the estimate of $d(t_{k-1})$ is 
\begin{align}
\hat{d}(t_{k-1}) = E_d^{-1}(t_{k-1}) C_k^{-1} (y_k - C_k {x}^*_{k|k-1}).
\end{align}

Substituting the above equation to \eqref{e:x_est_special}, it follows
\begin{align}
\hat{x}_{k|k} =& {x}^*_{k|k-1} + L_k (y_k - C_k {x}^*_{k|k-1})
\end{align}
where 
\begin{align}
L_k &= K_k + (I - K_k C_k) E_d(t_{k-1}) E_d^{-1}(t_{k-1}) C_k^{-1} \\
&= C_k^{-1}
\end{align}
It is found out that $K_k$ has no effect on the final estimate. This also explains why the optimal Kalman filter gain $K_k$ is not unique, as stated in \cite{Gillijns2007} which considers discrete time systems. 

Consequently, the following is obtained:
\begin{align}
\hat{x}_{k|k} &= {x}^*_{k|k-1} + C_k^{-1}y_k - {x}^*_{k|k-1}\\
&=C_k^{-1}y_k
\end{align}
This completes the proof.
\rlap{$\qquad \Box$}
\end{pf}

The stability of this simplified estimator is discussed below.

\begin{cor}
\label{thm:nd=ny}
The estimator denoted by \eqref{e:x_pred nod}-\eqref{e:x_est}, which is equivalent to \eqref{e:x_pred nod}, \eqref{e:d_est} and \eqref{e:x_est_one_step}, is a guaranteed stable estimator when $n_x=n_y=n_d$. Furthermore, it is robust with respect to initial condition errors. 
\end{cor}
\begin{pf}
Since $F_d(t_{k-1})=E_d^{-1}(t_{k-1}) C_k^{-1}$, $\tilde{A}_{k-1}$, $\tilde{G}_{k-1}$ and $\tilde{D}_k$ defined in \eqref{e:A tilde}, \eqref{e:G tilde} and \eqref{e:D tilde} is reduced to
\begin{align}
\tilde{A}_{k-1} =O, \tilde{G}_{k-1}=O, \tilde{D}_k = -C_k^{-1}.
\end{align}

The error dynamics of the estimator denoted by \eqref{e:x_pred nod}-\eqref{e:x_est} is reduced to
\begin{align}
e_{k|k} &= -C_k^{-1} v_k
\end{align}
Therefore, the estimator is always a stable estimator.

As it always only use the measurement, it is also free from the effect of initial condition errors.

This completes the proof.
\rlap{$ \quad \Box$}
\end{pf}

Accordingly, the following Corollary is also obtained:
\begin{cor}
\label{thm:nd=ny predictor}
The predictor denoted by \eqref{e:x_pred nod}-\eqref{e:x_pred d} is a guaranteed stable estimator when $n_x=n_y=n_d$. Furthermore, it is robust with respect to initial condition errors. 
\end{cor}
\begin{pf}
According to the proof of Theorem~\ref{thm:one-step}, the optimal Kalman gain has no effect on the final estimate. By setting $K_k=O$, then the estimator denoted by \eqref{e:x_pred nod}-\eqref{e:x_est} is equivalent to the predictor denoted by \eqref{e:x_pred nod}-\eqref{e:x_pred d}. Therefore, it follows from Corollary~\ref{thm:nd=ny} that this predictor is also always stable.

This can also be proved by using Theorem~\ref{thm:unbiased estimator stability}. As $F_d(t_{k-1})=E_d^{-1}(t_{k-1}) C_k^{-1}$, therefore
\begin{align}
\bar{A}_{k-1} &= ( I-E_d(t_{k-1})E_d^{-1}(t_{k-1}) C_k^{-1} C_k ) A_d(t_{k-1}) \\
&= O
\end{align}
Therefore, all the eigenvalues of $\bar{A}_{k-1}$ are within the unit circle. The predictor is always stable.
\rlap{$ \quad \Box$}
\end{pf}

\begin{remark}
The one-step Kalman filter proposed in this section, although proposed for continuous systems, is also applicable to discrete-time systems. Therefore, the following corollary is obtained.
\end{remark}

\begin{cor}
\label{thm:one-step discrete}
The estimator proposed in \cite{Gillijns2007c} is a guaranteed stable estimator when $n_x=n_y=n_d$. Furthermore, it is robust with respect to initial condition errors. 
\end{cor}

\section{Extension to design an unknown input observer}
\label{s:5}
The proposed estimator, denoted by \eqref{e:x_pred nod}-\eqref{e:x_est}, can be directly used to design an unknown input observer. The difference between an unknown input Kalman filter and an unknown input observer is that observer does not consider the noise effect. However, if the unknown input observer is designed in the same way as follows, it will be similar to an unknown input Kalman filter.

To differentiate from the proposed estimator, different variables are used. The unknown input observer is designed as follows:
\begin{align}
\dot{w} &= A \hat{x}, \text{ with } w(0)=\hat{x}(0) \\
\hat{d} &= F_d (y-Cw) \\
\dot{z} &= A \hat{x} + E \hat{d}, \text{ with } z(0)=\hat{x}(0) \\
\hat{x} &= z +L(y-Cz)
\end{align}
where $\hat{x}$ is the final state estimate and $\hat{d}$ is the unknown input estimate. $w$ and $z$ are intermediate variables and their initial values are $\hat{x}(0)$. 

$L$ is the gain of the observer. $L$ can be designed to place the eigenvalues at desired locations. 
Note that the above unknown input observer is for time-invariant systems. For time-varying systems, then the design is the same as the \ac{R4SKF}. The asymptotic stability of this observer is similar to the estimator analyzed in Sec~\ref{s:3} and is therefore not repeated here. Note that the observer does not require to compute the error covariance.

When $n_d=n_y=n_x$, according to Theorem~\ref{thm:one-step}, the optimal Kalman gain $K_k$ has no effect on the final estimation. By selecting $L=C_k^{-1}$, the unknown input observer is exactly equivalent to the unknown input Kalman filter (\ac{R4SKF}) and has guaranteed stability.

%
%
%

Note that for the special case when $n_d=n_y$, a stable unknown input observer designed by Saif and Guan \cite{Saif1993} may not exist. However, the estimator proposed in this paper, denoted by \eqref{e:x_pred nod}-\eqref{e:x_est}, can guarantee its stability according to Corollary~\ref{thm:nd=ny}.

\section{State and unknown input estimation using augmented Kalman filters}
\label{s:6}
So far the model of unknown inputs are not required when dealing with simultaneous state and unknown input estimation. The other type of Kalman filter for unknown input estimation, augmented state Kalman filter, requires the modeling of unknown inputs. This section will introduce the \ac{A2KF} and then compare its estimation performance with the \ac{R4SKF}.

\subsection{Adaptive augmented Kalman filter}
\label{s:61}
As the model of unknown inputs are unknown, the following stochastic process is usually used instead:
\begin{align}
\dot{d}(t) = w^d(t)
\end{align}
where $w^d(t) \sim N(0,Q^d(t))$. By augmenting the unknown inputs in the state vector, the following model is obtained:
\begin{align}
\dot{x}^a(t) & = A^a(t) x^a(t) + B^a(t) u(t) + G^a(t) w^a(t) \\
y_{k} & = C^a_{k} x^a_{k}  + \nu_{k}
\end{align}
where
\begin{align}
x^a &= \begin{bmatrix}
x \\ d
\end{bmatrix},
A^a = \begin{bmatrix}
A & E \\O & O
\end{bmatrix},
B^a = \begin{bmatrix}
B \\ O
\end{bmatrix},\\
w^a &= \begin{bmatrix}
w \\ w^d
\end{bmatrix},
G^a = \begin{bmatrix}
G & O \\O & I
\end{bmatrix},
C^a = \begin{bmatrix}
C & O 
\end{bmatrix}
\end{align}
Since the original process noise $w(t)$ is augmented with $w^d(t)$, the process noise covariance matrix for the augmented Kalman filter is $Q^a(t)=\begin{bmatrix}
Q(t) & \\ & Q^d(t)
\end{bmatrix}$. 
By assuming the prior knowledge of $Q^d(t)$, the state and unknown input can be estimated using the standard Kalman filter. Increase of the state dimension could result in heavier computational load. Therefore, two-stage Kalman filters are proposed which split the augmented filter into a bias-free filter and a bias filter \cite{Friedland1969,Hsieh1999,Lu2015e}. 

However, $Q^d(t)$ is still unknown. To address that,  Lu et al. \cite{Lu2016} propose an adaptive covariance method to estimate $Q^d$ in real time. An extension of this adaptive covariance estimation of unknown inputs to nonlinear systems is performed in \cite{Lu2019}. In this paper, the augmented Kalman filter is extended with the adaptive estimation of $Q^d(t)$. For the details of the derivation of $Q^d(t)$, please refer to \cite{Lu2016,Lu2019}. Here, only the results are shown in the following:

Define 
\begin{align}
\label{e:C_gamma_0}
C^{\gamma}_0=C^{\gamma}_k - C_k G(t_{k-1}) Q(t_{k-1}) G^T(t_{k-1}) C_k^T \Delta t - R_k,
\end{align}
then the estimation of $Q^d(t_k)$ can be solved using
\begin{align}
C_k E_d(t_{k-1}) Q^d(t_{k-1}) E_d^T(t_{k-1}) C_k^T = C^{\gamma}_0
\end{align}
where $C^{\gamma}_k$ is the actual innovation covariance \cite{Lu2016,Lu2019}.
Note that in \cite{Lu2016,Lu2019}, $C_k E_d(t_{k-1})$ is invertible. However, in this paper, $C_k E_d(t_{k-1})$ may not be invertible. Consequently, the estimation of $Q^d(t_{k-1})$ is given as follows:
\begin{align}
\label{e:Qd_est}
(C_k E_d(t_{k-1}) )^{+} C^{\gamma}_0 (E_d^T(t_{k-1}) C_k^T)^{+}
\end{align}
where $(C_k E_d(t_{k-1}) )^{+}$ denotes the Moore-Penrose inverse. $E_d(t_{k-1})=E(t_{k-1})\Delta t$ where $\Delta t$ is the sampling time and can also be considered as a scaling factor. In case there are negative numbers in $C^{\gamma}_0$, then only the main diagonal elements are used.

\subsection{Comparison between the \ac{A2KF} and the \ac{R4SKF}}
\label{s:62}
In this subsection, the unknown input estimation performance of the \ac{R4SKF} and the \ac{A2KF} is analyzed and compared. For estimation using Kalman filters, the covariance of estimation error is of key concern. Therefore, we will compare the error covariance of their unknown input estimates.

Using \eqref{e:ed}, it is possible to compute the covariance of the unknown input estimation error as follows:
\begin{align}
P^d(t_{k-1}) =& E[e^d(t_{k-1}) (e^d(t_{k-1}))^T] \\
= & F_d(t_{k-1})[ C_k A_d(t_{k-1}) P_{k-1|k-1} A_d^T(t_{k-1}) C_k^T \nonumber \\
 & + C_k G(t_{k-1})Q(t_{k-1}) G^T(t_{k-1})C_k^T \Delta t \nonumber \\
 & +R_k ] F_d^T(t_{k-1})
\end{align}

In steady state, $P_{k-1|k-1}$ is of small magnitudes. Therefore, the estimation error covariance of the unknown input reduces to
\begin{align}
\label{e:Pd_simplified}
P^d(t_{k-1}) =& F_d(t_{k-1})[ C_k G(t_{k-1})Q(t_{k-1}) G^T(t_{k-1})C_k^T \Delta t \nonumber \\
 & +R_k ] F_d^T(t_{k-1})
\end{align}

Recall that $F_d(t_{k-1})=(C_k E_d(t_{k-1}))^{+}$ where $E_d(t_{k-1})=E(t_{k-1}) \Delta t$. Without losing generality, let use assume $C_k=I$ and $E(t_{k-1})=I$  for this analysis. Consequently, the estimation error covariance is 
\begin{align}
\label{e:Pd_simplified2}
P^d(t_{k-1}) =& [ C_k G(t_{k-1})Q(t_{k-1}) G^T(t_{k-1})C_k^T \Delta t \nonumber \\
& +R_k ]/ (\Delta t^2)
\end{align}
It is easy to observe that the measurement noise covariance is magnified with $1/\Delta t^2$. This is equivalent to differentiating the measurement noise as performed in \cite{Yong2017}. Consequently, The estimation error is largely dependent on the magnified measurement noise. 

On the contrary, determining the error covariance matrix of the augmented Kalman filters is a bit tricky. In short, it depends on the selected $Q^d$. For the \ac{A2KF}, $Q^d$ is estimated in real time using \eqref{e:Qd_est}. It has a similar form as \eqref{e:Pd_simplified}. The main difference is that the value in \eqref{e:Pd_simplified} is fixed whereas \eqref{e:Qd_est} is variant depending on $C^{\gamma}_0$ which is defined in \eqref{e:C_gamma_0}. 

\begin{figure}
\includegraphics[width = 0.5\textwidth]{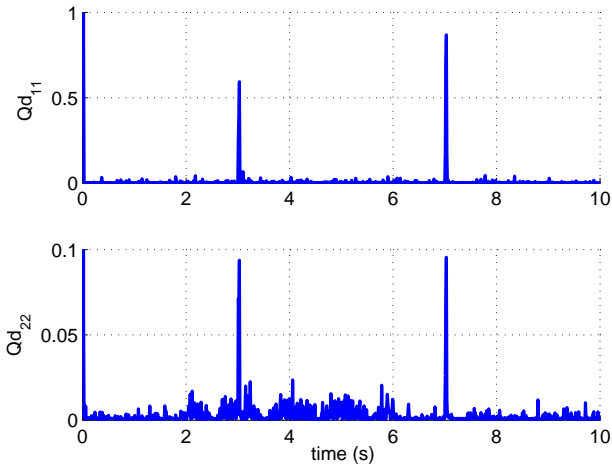}
\caption{The main diagonal elements of $Q^d$ estimated by the \ac{A2KF}.}
\label{f:Qd_est}
\end{figure}

Apparently, if the unknown input is zero,  $C^{\gamma}_0$ will also be zero-mean, which results in a $Q^d$ with a small magnitude. If the unknown input is non-zero, $Q^d$ is proportional to the derivative of the unknown input. For instance, if the unknown input is a step signal. The magnitude of $Q^d$ will significantly increase at the step time and then it will quickly decrease to a small value. This will also be seen in Fig.~\ref{f:Qd_est} in Sec.~\ref{s:7}.

\section{Illustrative examples}
\label{s:7}
The \ac{R4SKF} and the \ac{A2KF} will both be implemented to estimate the state and unknown input. Their estimation performances will be compared in different cases.

The system matrices are as follows:
\begin{align}
A(t)  & =  \begin{bmatrix}
	1.9527 & -0.0075 &  0.0663 &  0.0437 \\
    0.0017 & 1.0452 &   0.0056 &  -0.0242 \\
    0.0092 & 0.0064 &  -0.1975 &  0.00128 \\
    	0 &		0   &		1 &		0
\end{bmatrix}, \nonumber 
\end{align}
\begin{align}
B(t) & =  \begin{bmatrix}
	0.554  & 0.156 \\
    0.246  & -0.982 \\
    0.320  &  0.560 \\
       0   &  0 
\end{bmatrix}, \ \  \indent  E(t) = B(t), \nonumber  
\end{align}
\begin{align}
G(t)&= \begin{bmatrix}
	1 & 0 &  0 &  0 \\
    0 & 1 &   0 &  0 \\
    0 & 0 &  1 &  0 \\
    0 &	0 &  0 &  1
\end{bmatrix} , \ \ 
C_k  = \begin{bmatrix}
	1  & 0  & 0  & 0 \\
    0  & 1  & 0  & 0 \\
    0  & 0  & 0  & 1
\end{bmatrix}, \nonumber 
\end{align}
\begin{align}
Q(t) &= \text{diag}(10^{-6},10^{-6},10^{-6},10^{-6}) \nonumber \\
R_k &= \text{diag}(10^{-7},10^{-7},10^{-7} ) \nonumber \\
u(t) &= [0,0]^T \nonumber
\end{align}

The true unknown inputs are:
\begin{align}
d(t) &= \begin{bmatrix}
d_1(t) \\ d_2(t)
\end{bmatrix} \\
d_1(t) &= \begin{cases}
0.5, \quad  3<t\leq 7 \\
0, \quad\ \ \  \text{otherwise}
\end{cases}, \\
\label{e:d2}
d_2(t) & = \begin{cases}
0.4 \sin (2 \pi f_0 (t-2)), \quad  2<t\leq 6 \\
0, \quad \quad   \text{otherwise}
\end{cases},
\end{align}
where $f_0$ is the frequency of the unknown input. 


The true initial condition is
\begin{align}
 x_{0}=[0,0,0,0]^T 
\end{align}
The initial condition for the \ac{R4SKF} and \ac{A2KF} are both set to the following:
\begin{align}
 \hat{x}_{0|0}=[10,10,10,10]^T. 
\end{align}
Consequently, there are initial condition errors.

\begin{figure}
\subfigure[Unknown input estimation using the \ac{R4SKF}, case 1]{
\includegraphics[width = 0.5\textwidth]{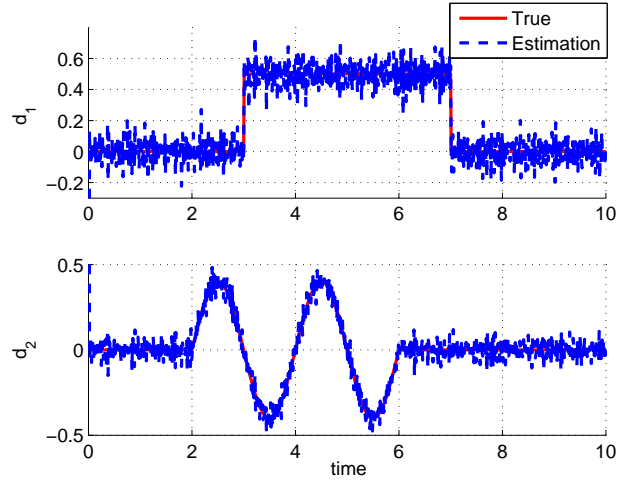}
\label{f:d_est_R4SKF}
}
\subfigure[Unknown input estimation using the \ac{A2KF}, case 1.]{
\includegraphics[width = 0.5\textwidth]{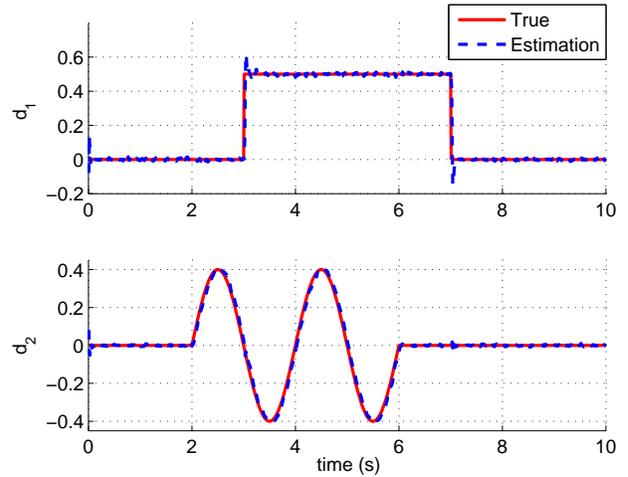}
\label{f:d_est_A2KF}
}
\caption{Unknown input estimation using the two approaches, case 1}
\label{f:d_est}
\end{figure}

\subsection{Case 1: an unstable fourth-order system}
In this case, $f_0$, which is the frequency of the sine function defined in \eqref{e:d2}, is set to 0.5. The two approaches are both implemented.

The unknown input estimation results using the \ac{R4SKF} and \ac{A2KF} are displayed in Figs.~\ref{f:d_est_R4SKF} and \ref{f:d_est_A2KF}, respectively. It is seen that both approaches can achieve unbiased estimation. However, the estimation results using the \ac{R4SKF} is noisier than the one using the \ac{A2KF}.

The \ac{RMSE}s of the state and unknown input estimation using two approaches are presented in Table~\ref{t:RMSE}. The minimum \ac{RMSE}s are highlighted in bold. As seen, the \ac{A2KF} achieves the minimum \ac{RMSE}s in all the estimates. The \ac{RMSE}s for state estimation using the two approaches are similar. However, the \ac{RMSE}s for the unknown input estimation using the \ac{R4SKF} is almost three times the one obtained using the \ac{A2KF}.

The estimation of $Q^d$ using the \ac{A2KF} is shown in Fig.~\ref{f:Qd_est}. As stated in Sec.~\ref{s:62}, $Q^d$ will be of small magnitudes when the unknown input is zero and proportional to the derivative of the unknown input. As $d_1$ is of step-type. The magnitudes of the diagonal elements of $\hat{Q}^d$ both increase to a large value when $t=3$ and $t=7$ and then decrease to a small value immediately. For the second main diagonal element of $\hat{Q}^d$, its magnitude is also large during $2<t<6$ s as caused by $d_2$.

\begin{table*}
\setlength{\tabcolsep}{3pt} 
\renewcommand{\arraystretch}{1.5} 
\caption{RMSEs of the state and unknown input estimation}
\centering
\begin{tabular}{ cccccccc }
\hline \hline
   & Methods & $x_1$ & $x_2$ & $x_3$ & $x_4$  & $d_1$ & $d_2$ \\ \hline
\multirow{2}{*}{Case 1}
 & \ac{R4SKF}   & 0.000315  & 0.000302   & 0.000360   &  0.000122  & 0.075171   & 0.044637   \\
 & \ac{A2KF}	& \textbf{0.000248}  & \textbf{0.000265}   & \textbf{0.000323}   &  \textbf{0.000122 } & \textbf{0.029995}   & \textbf{0.015039}	\\
  \hline 
\multirow{2}{*}{Case 2}
 & \ac{R4SKF}  & 0.000315  & \textbf{0.000302}   & \textbf{0.000359}   &  \textbf{0.000122} &  0.075171   & \textbf{0.044637}   \\
 & \ac{A2KF}	& \textbf{0.000287}  & 0.000645   & 0.000500   &  \textbf{0.000122} &  \textbf{0.029689}   & 0.068015	\\
  \hline 
 \multirow{2}{*}{Case 2}
 & \ac{R4SKF}  & 0.003146  & 0.003016   & 0.003128   &  0.000474 &  0.751464   & 0.446227   \\
 & \ac{A2KF}	& \textbf{0.001767}  & \textbf{0.001729}   & \textbf{0.002047}   &  \textbf{0.000472} &  \textbf{0.069634}  & \textbf{0.039664}	\\
  \hline   \hline
\end{tabular}
\label{t:RMSE}
\end{table*}

\begin{figure}
\subfigure[Unknown input estimation using the \ac{R4SKF}, case 2]{
\includegraphics[width = 0.5\textwidth]{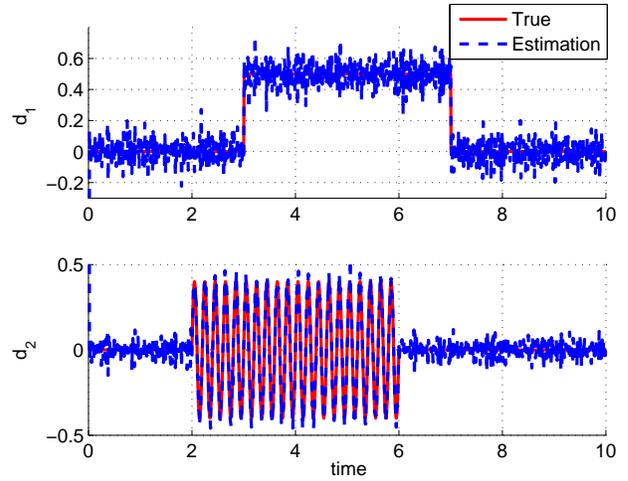}
\label{f:d_est_R4SKF_faster}
}
\subfigure[Unknown input estimation using the \ac{A2KF}, case 2.]{
\includegraphics[width = 0.5\textwidth]{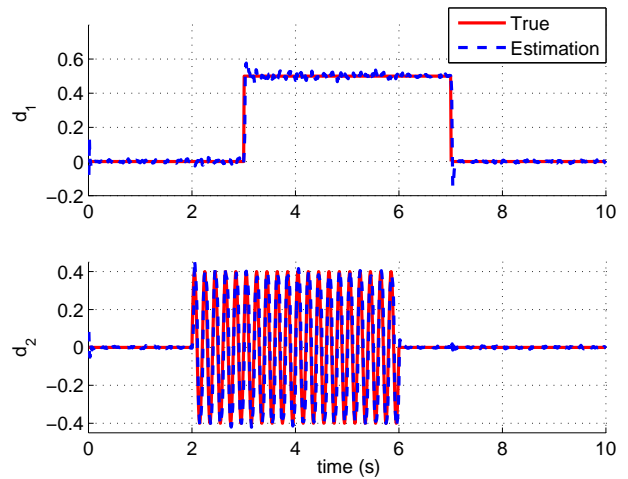}
\label{f:d_est_A2KF_faster}
}
\caption{Unknown input estimation using the two approaches, case 2.}
\label{f:d_est_faster}
\end{figure}

\subsection{Case 2: Faster unknown inputs}
In this case, faster unknown inputs are considered. The frequency of the sine function defined in \eqref{e:d2} is set to 5, which is ten times the one in case 1. This is to validate the estimation performance of the two approaches when addressing highly dynamic unknown inputs.

The unknown input estimation results using the \ac{R4SKF} and \ac{A2KF} are shown in Figs.~\ref{f:d_est_R4SKF_faster} and \ref{f:d_est_A2KF_faster}, respectively. The results are very similar to the one obtained in case 1. Both approaches can estimate the fast unknown inputs satisfactorily. 

The \ac{RMSE}s of the state and unknown input estimation using two approaches are given in Table~\ref{t:RMSE}. The minimum \ac{RMSE}s are highlighted in bold. In this case, the approaches behave equally well. The estimation of $d_2$ using the \ac{A2KF} is a bit worse than the one using the \ac{R4SKF}. The performance of the \ac{A2KF} can be readily increased by scaling the estimated $Q^d$ with $1/\Delta t$. However, this is out of the scope of this paper.

\begin{figure}
\subfigure[Unknown input estimation using the \ac{R4SKF}, case 3.]{
\includegraphics[width = 0.5\textwidth]{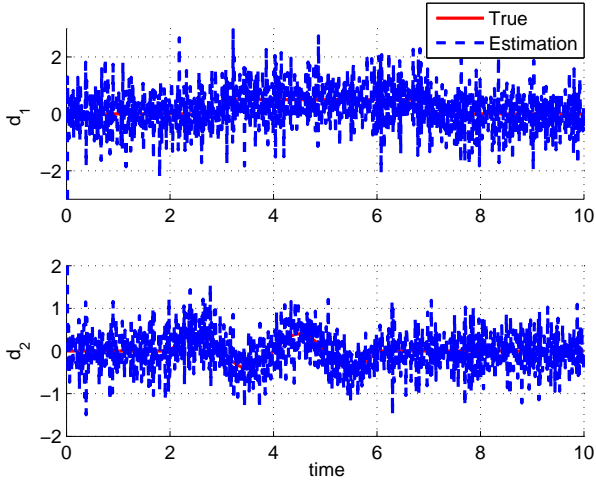}
\label{f:d_est_R4SKF_noisier}
}
\subfigure[Unknown input estimation using the \ac{A2KF}, case 3.]{
\includegraphics[width = 0.5\textwidth]{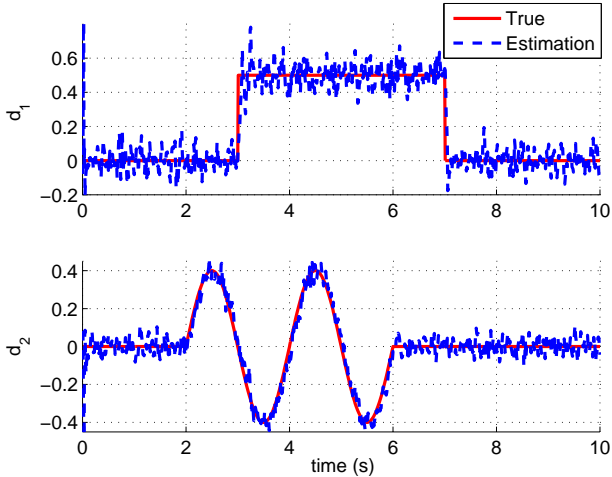}
\label{f:d_est_A2KF_noisier}
}
\caption{Unknown input estimation using the two approaches, case 3. Note the scale differences in (a) and (b).}
\label{f:d_est_noisier}
\end{figure}

\subsection{Case 3: Larger measurement noise}
In this case, the standard deviation of the measurement noise is increased 10 times, which results in the following $R_k$:
\begin{align}
\text{diag}(10^{-5},10^{-5},10^{-5} ) 
\end{align}
This is to validate the estimation performance of the two approaches in terms of different measurement noise. 

The unknown input estimation results using the \ac{R4SKF} and \ac{A2KF} are demonstrated in Figs.~\ref{f:d_est_R4SKF_noisier} and \ref{f:d_est_A2KF_noisier}, respectively. It is noted that the estimation performance of the two approaches both degrade when the measurements are noisier, which is expected. However, the \ac{R4SKF} is significantly more affected by the increased measurement noise. Note the scale differences in the two figures. It is difficult to tell the shape of the unknown inputs from Fig.~\ref{f:d_est_R4SKF_noisier} whereas the shape of the unknown inputs are estimated reasonably well in Fig.~\ref{f:d_est_A2KF_noisier}.

The \ac{RMSE}s of the two approaches can be found in Table~\ref{t:RMSE}. All the minimum \ac{RMSE}s are again obtained using the \ac{A2KF}. The \ac{RMSE}s of the state estimation using the two approaches are similar while the \ac{A2KF} behaves better. However, the \ac{RMSE}s of the unknown input estimation using the \ac{R4SKF} is ten times larger than the one obtained using the \ac{A2KF}, which demonstrates the superior estimation performance of the \ac{A2KF}. Increasing the process noise will result in similar results.

\acresetall
\section{Conclusion}
\label{s:8}
This paper considers the simultaneous state and unknown input estimation for continuous stochastic systems. A \ac{R4SKF} is proposed to estimate the state and unknown input in an unbiased sense. The stability of the proposed estimator is analyzed. For special cases when the number of unknown inputs are equal to that of the measurements, it is proved that the optimal Kalman gain has no effect on the final state and unknown input estimation. Furthermore, a one-step estimator is derived to estimate the states. It is proven that the estimator is always stable and robust to initial condition errors.

It was interesting to notice that the design of an unknown input observer can be similar as an unknown input Kalman filter for continuous systems. It is observed that when the number of unknown inputs is the same as that of the measurements, the two are exactly equivalent.

Finally, the other type of Kalman filters, the \ac{A2KF}, which can also address state and unknown input estimation is compared with the \ac{R4SKF} in terms of estimation error covariances. It is found out that the estimation error of the \ac{R4SKF} is largely dependent on the differentiated measurement noise whereas that of the \ac{A2KF} is not. Simulation examples demonstrate that the \ac{A2KF} is a better option for state and unknown input estimation of continuous stochastic systems.

%

\bibliographystyle{plain}        
\bibliography{lite_AFTC}    

\begin{thebibliography}{10}

\bibitem{Cheng2009}
Yue Cheng, Hao Ye, Yongqiang Wang, and Donghua Zhou.
\newblock {Unbiased Minimum-Variance State Estimation for Linear Systems with
  Unknown Input}.
\newblock {\em Automatica}, 45(2):485--491, February 2009.

\bibitem{Darouach1997}
M.~Darouach and M.~Zasadzinski.
\newblock {Unbiased Minimum Variance Estimation for Systems with Unknown
  Exogenous Inputs}.
\newblock {\em Automatica}, 33(4):717--719, 1997.

\bibitem{Darouach2003}
M.~Darouach, M.~Zasadzinski, and M.~Boutayeb.
\newblock {Extension of Minimum Variance Estimation for Systems with Unknown
  Inputs}.
\newblock {\em Automatica}, 39(5):867--876, May 2003.

\bibitem{Ducard2008}
Guillaume Ducard and Hans~Peter Geering.
\newblock {Efficient Nonlinear Actuator Fault Detection and Isolation System
  for Unmanned Aerial Vehicles}.
\newblock {\em Journal of Guidance, Control, and Dynamics}, 31(1):225--237,
  2008.

\bibitem{Friedland1969}
Bernard Friedland.
\newblock {Treatment of Bias in Recursive Filtering}.
\newblock {\em IEEE Transactions on Automatic Control}, 14(4):359--367, 1969.

\bibitem{Gillijns2007c}
Steven Gillijns and Bart {De Moor}.
\newblock {Unbiased Minimum-Variance Input and State Estimation for Linear
  Discrete-Time Systems}.
\newblock {\em Automatica}, 43(1):111--116, January 2007.

\bibitem{Gillijns2007}
Steven Gillijns and Bart {De Moor}.
\newblock {Unbiased Minimum-Variance Input and State Estimation for Linear
  Discrete-Time Systems with Direct Feedthrough}.
\newblock {\em Automatica}, 43(5):934--937, May 2007.

\bibitem{Hou1998}
M.~Hou and R.~J. Patton.
\newblock {Optimal Filtering for Systems with Unknown Inputs}.
\newblock {\em IEEE Transactions on Automatic Control}, 43(3):445--449, 1998.

\bibitem{Hsieh2000}
Chien-Shu Hsieh.
\newblock {Robust Two-Stage Kalman Filters for Systems with Unknown Inputs}.
\newblock {\em IEEE Transactions on Automatic Control}, 45(12):2374--2378,
  2000.

\bibitem{Hsieh2009}
Chien-Shu Hsieh.
\newblock {Extension of unbiased minimum-variance input and state estimation
  for systems with unknown inputs}.
\newblock {\em Automatica}, 45(9):2149--2153, September 2009.

\bibitem{Hsieh2020}
Chien-Shu Hsieh.
\newblock Time-distributed multi-step delayed input and state estimation.
\newblock {\em Automatica}, 112:108700, 2020.

\bibitem{Hsieh1999}
Chien-Shu Hsieh and Fu-Guang Chen.
\newblock {Optimal Solution of the Two-Stage Kalman Estimator}.
\newblock {\em IEEE Transactions on Automatic Control}, 44(1):194--199, 1999.

\bibitem{Kim2020}
Hunmin Kim, Pinyao Guo, Minghui Zhu, and Peng Liu.
\newblock Simultaneous input and state estimation for stochastic nonlinear
  systems with additive unknown inputs.
\newblock {\em Automatica}, 111:108588, 2020.

\bibitem{Kitandis1987}
Peter~K. Kitanidis.
\newblock {Unbiased Minimum-variance Linear State Estimation}.
\newblock {\em Automatica}, 23(6):775--778, 1987.

\bibitem{Lu2019}
P.~{Lu}, T.~{Sandy}, and J.~{Buchli}.
\newblock {Adaptive Unscented Kalman Filter-based Disturbance Rejection With
  Application to High Precision Hydraulic Robotic Control}.
\newblock In {\em IEEE/RSJ International Conference on Intelligent Robots and
  Systems (IROS 2019)}, pages 4365--4372, Macau, China, 2019.

\bibitem{Lu2015g}
P.~Lu, L.~{Van Eykeren}, E.~van Kampen, and Q.~P. Chu.
\newblock {Selective-Reinitialisation Multiple Model Adaptive Estimation for
  Fault Detection and Diagnosis}.
\newblock {\em Journal of Guidance, Control, and Dynamics}, 38(8):1409--1425,
  2015.

\bibitem{Lu2015e}
P.~Lu, L.~{Van Eykeren}, E.~van Kampen, C.~C. de~Visser, and Q.~P. Chu.
\newblock {Aircraft Inertial Measurement Unit Fault Detection and Diagnosis
  with Application to Real Flight Data}.
\newblock {\em Journal of Guidance, Control, and Dynamics}, 38(12):2467--2475,
  2015.

\bibitem{Lu2015}
Peng Lu, Laurens {Van Eykeren}, E.~van Kampen, Cornelis~Coen de~Visser, and
  Q.~P. Chu.
\newblock {Double-Model Adaptive Fault Detection and Diagnosis Applied to Real
  Flight Data}.
\newblock {\em Control Engineering Practice}, 36:39--57, March 2015.

\bibitem{Lu2016a}
Peng Lu, E.~van Kampen, Cornelis de~Visser, and Q.~P. Chu.
\newblock {Framework for state and unknown input estimation of linear
  time-varying systems}.
\newblock {\em Automatica}, pages 145--154, 2016.

\bibitem{Lu2016}
Peng Lu, E.~van Kampen, Cornelis de~Visser, and Q.~P. Chu.
\newblock {Nonlinear Aircraft Sensor Fault Reconstruction in the Presence of
  Disturbances Validated by Real Flight Data}.
\newblock {\em Control Engineering Practice}, 49:112--128, 2016.

\bibitem{Saif1993}
M.~{Saif} and Y.~{Guan}.
\newblock A new approach to robust fault detection and identification.
\newblock {\em IEEE Transactions on Aerospace and Electronic Systems},
  29(3):685--695, July 1993.

\bibitem{Sarkka2007}
Simo S\"{a}rkk\"{a}.
\newblock {On Unscented Kalman Filtering for State Estimation of
  Continuous-Time Nonlinear Systems}.
\newblock {\em IEEE Transactions on Automatic Control}, 52(9):1631--1641, 2007.

\bibitem{Yong2017}
S.~Z. {Yong}, M.~{Zhu}, and E.~{Frazzoli}.
\newblock Simultaneous input and state estimation for linear time-varying
  continuous-time stochastic systems.
\newblock {\em IEEE Transactions on Automatic Control}, 62(5):2531--2538, May
  2017.

\bibitem{Yong2016}
Sze~Zheng Yong, Minghui Zhu, and Emilio Frazzoli.
\newblock A unified filter for simultaneous input and state estimation of
  linear discrete-time stochastic systems.
\newblock {\em Automatica}, 63:321 -- 329, 2016.

\end{thebibliography}

\end{document}